# A Survey of Molecular Hydrogen in the Crab Nebula

E. D. Loh, J. A. Baldwin[1], Z. K. Curtis,

Department of Physics and Astronomy, Michigan State University, East Lansing, MI 48824-2320 USA

(loh@msu.edu, baldwin@pa.msu.edu, curtisza@msu.edu)

G. J. Ferland

Department of Physics, University of Kentucky, Lexington, KY 40506, USA
(gary@pa.uky.edu)

C. R. O'Dell

Physics & Astronomy Department, Vanderbilt University, Nashville, Tennessee 37240, USA
(cr.odell@vanderbilt.edu)

A. C. Fabian

Institute of Astronomy, University of Cambridge, Madingley Road, Cambridge CB3 0HA, UK (acf@ast.cam.ac.uk)

and

Philippe Salomé
LERMA & UMR8112 du CNRS, Observatoire de Paris, 61 Av. de l'Observatoire, F-75014 Paris, France (philippe.salome@obspm.fr)

## Abstract

We have carried out a near-infrared, narrow-band imaging survey of the Crab Nebula, in the $H_2$ 2.12 μm and Brγ 2.17 μm lines, using the Spartan Infrared camera on the SOAR Telescope. Over a 2.8' × 5.1' area that encompasses about 2/3 of the full visible extent of the Crab, we detect 55 knots that emit strongly in the $H_2$ line. We catalog the observed properties of these knots. We show that they are in or next to the filaments that are seen in optical-passband emission lines. Comparison to HST [S II] and [O III] images shows that the $H_2$ knots are strongly associated with compact regions of low-ionization gas. We also find evidence of many additional, fainter $H_2$ features, both discrete knots and long streamers following gas that emits strongly in [S II]. A pixel-by-pixel analysis shows that about 6 percent of the Crab's projected surface area has significant $H_2$ emission that correlates with [S II] emission. We measured radial velocities of the [S II] λ6716 emission lines from 47 of the cataloged knots and find that most are on the far

---

(receding) side of the nebula. We also detect Brγ emission. It is right at the limit of our survey, and our Brγ filter cuts off part of the expected velocity range. But clearly the Brγ emission has a quite different morphology than the $H_2$ knots, following the long linear filaments that are seen in Hα and in [O III] optical emission lines.

*Key Words:* ISM: individual objects (Crab Nebula, M1) – ISM:supernova remnants

## *1. Introduction*

A characteristic feature of the Crab Nebula is the complex system of filaments that have formed in and around its interior synchrotron-emitting bubble. At a distance of only 2 kpc (Trimble, 1973), the Crab offers the unique opportunity to examine the properties of individual filaments with high spatial resolution (1" = 0.01 pc). This makes the Crab our best available laboratory for studying the behavior of gas condensations in a hostile environment of high-energy radiation and particles.

The Crab filaments have been the subject of many previous investigations (see the review articles by Davidson & Fesen 1985 and by Hester 2008). Most of these concentrated on the optical emission lines from the ionized gas, generally with the aim of measuring the chemical abundances in the aftermath of a supernova explosion (Davidson 1973, 1978: Henry & MacAlpine 1982; Pequignot & Dennefeld 1983; MacAlpine et al. 1996; MacAlpine & Satterfield 2008). The Crab's synchrotron radiation field has been taken into account in this work, but generally not the effects of the high-energy particles, which must also be present.

Molecular gas also exists within at least some of the filaments. In addition, dusty condensations can be seen in silhouette against background synchrotron emission (Fesen & Blair 1990; Sankrit et al. 1998). More directly, Graham et al. (1990; hereafter G90) detected near-IR emission from the $H_2$ 2.12 μm line at two of the three locations that they observed within the Crab. This line presumably comes from dense molecular cores that emit strongly from at least some filaments.

The spectra of these dense, cool regions can tell a much fuller story about the ways in which X-rays, γ-rays and high-energy particles affect the physical conditions within the filaments. G90 gave a prescient discussion of the possible roles of UV fluorescence, collisional excitation, and cosmic-ray heating. They concluded that the observed $H_2$ emission is likely produced by heating due to some combination of shocks driven by ionization heating and penetrating relativistic particles from the surrounding synchrotron emitting plasma.

Here we present results from our new near-IR narrow-band imaging survey in the $H_2$ 2.12 μm and Brγ 2.17 μm emission lines. The G90 measurements of $H_2$ in the Crab were made through a 19" diameter aperture, which does not resolve the complicated filamentary structure of the Crab. In addition, they only looked at three locations in the Crab. Our observations cover 2/3 of the full face of the Crab Nebula at subarcsec resolution and reach much fainter surface brightness limits than the G90 measurements. We have already reported the discovery of one bright, compact $H_2$ knot as part of this survey (Loh et al. 2010). As will be described below, we find at least 55 molecular cores or emission knots that are scattered along much of the crab's system of filaments. We will compare the morphology of these molecular knots to the structure visible in archival HST optical-passband narrow-band images. We also will briefly present results of follow-up optical spectroscopy aimed at measuring the radial velocities of individual knots in order to learn more about their locations in the three-dimensional nebula.

A companion paper (Loh et al. 2011) describes near-infrared spectroscopy of seven of these knots. Further papers, currently in preparation, will then discuss in more detail the physical nature of the observable knots, their relationship to other components of the filamentary gas, and their bearing on big-picture questions.

## 2. The near-infrared imaging survey

### 2.1 Observing procedure

To measure the 2.12μm 1–0 S(1) line of $H_2$ and the Brγ line, we observed with the Spartan Infrared Camera (Loh et al. 2004) on the SOAR Telescope with the $H_2$ (λ = 2.116μm, full width dλ = 0.031 μm), Cont3 (λ = 2.208 μm, dλ = 0.031 μm), and Brγ (λ = 2.162 nm, dλ = 0.021μm) filters. The Cont3 filter measured the synchrotron radiation. We used the wide-field mode, which has a plate scale of 68 mas/pixel. The field of the four detectors is 5.17' × 5.12' with 0.47' gaps.

We defined target A so that detectors 2 and 3 are on the Crab and target B so that detectors 0 and 1 are on the Crab. Targets A' and B' are offset by 40" in the northwest direction to cover the gap between detectors 0 and 1 and between detectors 2 and 3 (Figure 1). For each target, the off-Crab detectors measure the sky. However, the angular separation on the sky between detectors is not large enough to avoid the Crab completely, which can be seen in the figure, and this was corrected during data reduction. The observing sequence was targets A, A', B, and B' with an additional random dither of 5" before each exposure. This sequence was done with the $H_2$ filter and then with the Cont 3 filter and the Brγ filter. The exposure times in each filter, per exposure were 180 s ($H_2$), 180 s (Cont3) and 250 s (Brγ).

Table 1 is a summary of the observations. The quality of the images from the nights of 24 and 25 Dec 2009 are excellent. The image width of the composite of all images from these nights is 770 mas (FWHM). On 04 Jan, 07 Feb, and 24 Mar 2010, the seeing was much poorer, the wind was so strong that the images were elongated, or the observations were short. The observations from these nights confirm the brighter knots of $H_2$ but do not improve the measurement of the flux significantly.

### 2.2 Data reduction steps

Data reduction was done primarily with IRAF[2] supplemented by routines written in *Mathematica* and Fortran. The detectors' sensitivities show an offset that is low on physical columns that are a multiple of 128. A physical column is in the direction that changes fastest as the detector is read out. The direction of the physical columns rotates 90° between quadrants of the detector. We wrote a *Mathematica* routine *fixDetectorTiles* to measure and correct the offset at the end of each tile of 128×1024 pixels.

The detectors have bad pixels, which are defined to be pixels with a response less than 50% of the average (0.04, 0.02, 3.3, and 0.9% of detectors 0–3), hot pixels, those having a flux greater than 0.5DU s$^{-1}$ (0.5-0.8% of the pixels), and those having an unusually high variance in short exposures with no light (0.17% of the pixels). Isolated bad pixels were fixed by replacing the value with the mean of the surrounding pixels. Clusters of bad pixels were flagged so that they were not used in creating the final mosaic of all of the images.

---

[2] IRAF is distributed by the National Optical Astronomy Observatory, which is operated by the Association of Universities for Research in Astronomy (AURA) under cooperative agreement with the National Science Foundation.

The instrument has a light leak for which the intensity is consistent with thermal radiation of something having the temperature of a sensor on the strut of the telescope. There is a leak between the inside of the instrument and the vacuum wall, which is at ambient temperature. There is a smaller leak through the wall that separates the halves of the instrument upstream and downstream from the Lyot stop. These two light leaks have a distinctive shape, the former in detector 1 and the latter in detector 2. The residual light leak was removed by scaling the light leak by hand to remove the distinctive shape.

The *Mathematica* routine *erodeCosmicRays* was written to remove cosmic rays. On the first pass, *erodeCosmicRays* removes the edge of a cosmic ray and on successive passes shrinks the region affected by the cosmic ray. We did not use the IRAF routines *cosmicrays* or *crnebula* because these removed the centers of stars. The rate was 4Hz for a single detector or a flux $700 s^{-1} m^{-2}$. (A cosmic ray covering multiple contiguous pixels counts as a single cosmic ray.) The intensity of a cosmic ray is proportional to the number of pixels covered.

We next formed a separate median sky image for each of the four detectors. The "sky" images are those taken with detectors 0 and 1 for target positions A and A' and those taken with detectors 2 and 3 for targets B and B' (these are the images that lie outside of the heavy dashed white rectangle on Fig. 1). Each sky image was scaled to match the median of the central 1024×1024 pixels of the first image used in the median image, and the images were aligned in pixel coordinates rather than in world coordinates. The resulting median sky image was then subtracted from each Crab image (*i.e.* from each image that is not a sky image) taken with the same detector.

The world coordinate system (WCS) of the images requires that the zero point be determined for each image, since the telescope coordinates vary by a few arcsec even when tracking a star. A test of a field in Carina showed that there is a non-zero distortion between the Spartan positions and those of the 2MASS catalog (Skrutskie et al 2006 ). The distortion in the corner of the detector where it is greatest is (95 ± 24,−46 ± 18) mas, which is (1.4 ± 0.7, 0.7 ± 0.3) pixel. Furthermore the zero point of the WCS determined by the IRAF routine msccmatch and by the DAOPhot positions disagree by 0.8 pixel (Loh & Abdul Rashid 2010).

The WCS header uses the Simple Polynomial (SIP) prescription for specifying distortion (using instrument coordinates), whereas the IRAF routines use the TNX specification (using sky coordinates). We wrote a *Mathematica* routine *writeTNXHeader* to translate the FITS header from SIP to TNX. The IRAF routine *msccmatch* was then used to determine the zero point and rotation of the telescope coordinates by comparing to positions of stars in the 2MASS catalogue. Thus our final images are on the J2000 coordinate system used by 2MASS.

The Crab images were then resampled onto a rectangular grid using the IRAF routine *mscimage*. The geometric distortion causes the size of a pixel to vary across the field by 6%. There is a circular pattern on each detector and a slope from detectors 0 and 3 to detectors 1 and 2. We chose to resample preserving brightness (DU $s^{-1}$ $sr^{-1}$) but not conserving flux (DU $s^{-1}$) so as not to have the detector edges appear in the combined image.

At this point, the general shape of the sky signal had been removed from each Crab image, but there was still an unknown intensity offset, since the sky varied by more than the brightness of the Crab nebula. Furthermore, the transparency of the sky may vary. We calibrated these effects with another *Mathematica* routine, *normalizeImages*, which uses the intensity of star-free regions and the photometry of stars in the image to determine the intensity offset and transparency (which are single numbers for each image). The sky brightness varied by 0.010 (the brightness of the sky is 0.2 in these units) or 5% of the sky, and the brightness for two detectors that were taken at the same time varied by 0.001 or 0.5% of the sky and by smaller amounts over shorter times. The brightness differences need not be zero, since all are compared with the first image taken with that combination of filter and detector. Each image was then corrected by both a multiplicative scale factor and an additive offset that were determined by this method.

The IRAF routine *imcombine* was then used to combine the roughly 130 individual Crab images in each filter to produce mosaic images of the entire Crab in each filter. Because of the contamination of some edges of our sky images by Crab emission, this was done in two steps. First, two separate mosaics of the Crab were made for each filter, one using just exposures with detectors 2 and 3 for which the telescope was pointed at targets A and A', and the other using just exposures with detectors 0 and 1 for which the telescope was pointed at targets B and B'. The regions on the Crab that were affected by the sky contamination were on the opposite edges of these two mosaics, so we could use the difference between the two sums to work out the amount of contamination in the sky images. This produced a correction image for each detector, which we subtracted from the individual geometrically-corrected images.

Next, we coadded the full image sets to produce final mosaic images in each filter. Then we subtracted the continuum image from the $H_2$ and $Br\gamma$ images, after scaling its intensity using the total counts measured in each mosaic in an area crossing the bright wisp that lies 8" NNW of the pulsar. Although at this stage the sky and Crab continuum had been properly removed, the actual flux scale was still not determined.

To derive the final flux calibration of the $H_2$ and $Br\gamma$ mosaic images, we used 2MASS K-band photometry of stars in the Crab field. This is preferable to using a separate standard star because it avoids having to measure the atmospheric extinction and possible nonlinearity in the detector response. A star was selected if it is single in our images and the 2MASS photometric quality flag for the K band is A or B, which indicates good accuracy. Furthermore, stars were rejected unless 0.22<J-H<0.54 and 0.03<H-K<0.21, to select F0-K2 dwarfs with moderate reddening and to eliminate objects with unusual colors (Covey et al. 2007).

Figure 2 shows the residuals from the conversion between $H_2$ instrumental magnitudes and K magnitudes. For the two cases where the absolute value of the residuals are greater than $2\sigma$ (at K=14.2 and 15.0), the residuals are similar for the $Br\gamma$ and C3 filters also, which indicates an outlier in the 2MASS measurement or variability. The error of the weighted mean for the $H_2$ filter is 0.003 mag, which is limited by the 2MASS photometry. For the $Br\gamma$ filter, the error of the weighted mean is 0.004 mag.

The absolute calibration was based on the flux of Vega determined by Campins, Rieke, & Lebofsky (1984). The conversion from Vega to a solar-type star and to the $H_2$ wavelength was based on the Kurucz' model for Vega with $T$ = 9400K, log $g$ = 3.95 (Kurucz 1979) at the two frequencies that straddle the K band and on the measurements of Labs & Neckel (1968) for the sun. The differential color shift between Vega and the sun for the K and $H_2$ filters is only 0.0006mag. It was assumed to be the same for all of the stars, since it is small.

Figure 3 shows the final products. The top panel shows our final co-added $H_2$ mosaic before subtracting the continuum, and the middle and lower panels show the $H_2$ and Brγ mosaics after the continuum subtraction. The region included in Fig. 3 was chosen to have relatively uniform signal-to-noise ratio, within the limits set by co-adding the complex pattern of individual images, and is the region used in the following analysis. The region is a 5.13'×2.82' rectangle at position angle 45° and centered at 11:38:03.3+22:01:01 (J2000).

The synchrotron continuum emission which provides the bright swirling background of the un-subtracted mosaic (Fig. 3a) cancelled out quite accurately. However, the continuum-subtracted mosaics did still show a number of low-level discontinuities running in the horizontal and vertical directions, caused by the edges of the individual images used in the sum. We removed them by applying very broad median filters running in just the vertical and horizontal directions. This procedure did not filter out the small fluctuations oriented at roughly 45 deg to the horizontal and vertical axes, so these show up as a faint herringbone pattern in the background of the continuum-subtracted mosaics in Figs. 3b and 3c. The continuum-subtracted mosaics shown in Fig. 3 have also had residual images from stars removed by interpolating over them, using the stars visible in the un-subtracted mosaic as a guide.

## *2.3 Position dependence of velocity coverage*

The filter response depends on the location of the object in the sky. The filters are placed near the Lyot stop, where the beam from a star is parallel. However, the incidence angle depends on the field position. Across the field, the incidence angle changes from normal to 7 deg (Stull 2009). Figure 4 shows the response vs. the Doppler shift of the $H_2$ and Brγ lines. The shift for a point at the edge of the field is about $-300$ km s$^{-1}$.

Figure 4 ignores the correction that is needed because the filter responses were measured at room temperature, rather than at 77K as used in the instrument. We do not have measurements of this effect for the filters actually used, but results for similar filters indicates wavelength shifts of about -80 km s$^{-1}$, which are not of major importance for interpreting the Crab observations.

The observed range of the Doppler velocities of [OIII] λ5007 is ±1600 km s$^{-1}$, and in the center of the nebula, where the projection effect is slight, the velocities are greater than 700 km s$^{-1}$or less than −700 km s$^{-1}$ (Fig. 2–4 of Clark et al. (1983)). Around the rim of the nebula, the Doppler velocities are low because the motion is mostly in the plane of the sky.

Brγ emitted by gas that is moving away from us with projected radial velocities greater than 600-800 km s$^{-1}$ will not be seen, because it falls outside the filter response; blueshifted Brγ is safely within the response of the filter. $H_2$ that is moving directly away will be seen but with poorer efficiency, while there is no problem with blueshifted $H_2$. Figure 4 shows these sensitivity cutoffs at positive velocities as contours over the same area of the sky that is covered by the images in Fig. 3, taking into account the pattern of telescope pointings shown in Fig. 1. The short-wavelength cutoffs are not shown because they do not affect the results for the Crab.

## *3. Archival HST images*

We also used archival HST WFPC2 images taken through the F502N and F673N filters, which respectively measure the [O III] λ5007 and the [S II] λ6717+λ6731 emission lines, although over fairly restricted velocity ranges. For each of these filters we assembled mosaic images that cover about 2/3 of the full Crab, using F673N images and F502N images taken at 5 separate pointings. We also constructed a mosaic image from WFPC2 images through the F547M continuum filter. Because the continuum is so small (1 percent or less), we did not remove it.

The Crab is expanding at a significant rate, so for each HST mosaic we only used images taken within two years of each other (in 1999 and 2000) so that the proper motions of the filaments and $H_2$ knots would not smear the images. We then adjusted the world coordinate system of each of the mosaic images by expanding the images around the center of the Crab's expansion at 05:34:32.94+22:00:50.1 (which uses the proper motion of Ng & Romani (2006) and our 2009 position of the pulsar) and assuming that the nebula has expanded uniformly from a point source with an effective starting date in the year 1176CE (which is the middle of the range of dates determined by extrapolating the Crab's current expansion rate backwards without allowing for its acceleration (Trimble 1968; Wyckoff & Murray 1977; Bietenholz et al. 1991)). We checked the accuracy of this expansion correction by repeating the procedure on additional HST images taken 6 years earlier, and found that some of the small features in the Crab filaments were misaligned between the two corrected images by 0.1–0.2" (1–2 HST pixels). This indicates that, in reality, the rate of expansion does vary somewhat with position, but not enough to strongly affect our comparison to the $H_2$ images.

### *4. Catalog of $H_2$ Knots*

We used the continuum-subtracted $H_2$ mosaic image (Fig 3b) to generate a catalogue of 55 $H_2$ - emitting knots (Table 2). To do this, we first visually examined the continuum-subtracted mosaic image and picked out extended (0.5" or larger) emission features whose average surface brightness exceeded the pixel-to-pixel background noise by an amount later measured to be about $1\sigma$. We verified that these were neither poorly subtracted stars nor just the residual edges of individual co-added images by blinking back and forth between the continuum-subtracted image and the input $H_2$ and continuum mosaic images that were used to make it. We then verified that we could see at least some sign of each of these emission features on separate continuum-subtracted $H_2$ mosaics made from two subsets of the full image set, one subset using the two nights in 2009 and the other using the three nights in 2010. This gave us further confidence that the $H_2$ emission features were not residual effects from the edges of individual images that went into the co-added sum, or from other problems that affected just one image. The knots selected in this way ranged from quite compact (but definitely non-stellar) clumps to quite extended features of roughly constant surface brightness.

We then smoothed the continuum-subtracted mosaic with a 9×9 boxcar, and drew an oversized box around each feature and also adjacent boxes around empty "sky" regions. Separately for each emission knot, we then computed the rms noise level in the sky regions, and flagged all pixels within the region of the emission knots that were more than $2\sigma$ above the mean sky level.

Finally, we went back to the unsmoothed continuum-subtracted mosaic and measured the flux above the mean sky level in each of the flagged pixels. This gave the total flux in the knot, except that a final correction of 15 percent or less (depending on the knot's radial velocity, discussed below) was made to account for the presence of two weak $H_2$ emission lines that our new infrared spectra (Loh et al. 2011) showed to be present in the continuum filter. The $H_2$ intensity ratios measured from the infrared spectra were used to determine this correction. The final corrected fluxes are listed in column 9 of Table 2, and the number of flagged pixels was converted into the total area in $arcsec^2$ covered by the knot (column 6 of the table). Dividing the flux by the surface area gave the mean surface brightness (column 7). We also measured the centroid right ascension and declination (in J2000 coordinates) listed in columns 2 and 3, and the peak surface brightness in the smoothed image within the area of the knot (column 8).

This pixel-flagging procedure produced a mask tracing the area covered by each $H_2$ knot. This mask was subsequently used to find the counts from the exact same areas on the sky in the Brγ image and also in the HST [S II] and [O III] images. The Brγ/$H_2$ flux ratio (after a correction of 15 percent or less to account for the additional weak $H_2$ lines that fall in the Brγ and continuum passbands) and the ratio of [S II]/[O III] count rates for each knot are listed in columns 10 and 11 of Table 2.

In this way we produced a catalogue of $H_2$ knots that covers a 2.8' × 5.1' area, about 2/3 of the full area of the optically-visible Crab Nebula. There are many additional weak, extended features visible in the continuum-subtracted $H_2$ mosaic, which we did not measure because they were blending into the residual correlated noise pattern left over from our various data reduction steps. Another artifact, the result of incomplete subtraction of the continuum of the wisps near the pulsar, can be seen 28" to the right of Knot 55 (in Fig. 7a, panel (b), described below). Figure 5 shows histograms of the mean and peak surface brightness in the knots. A careful visual inspection of the continuum-subtracted image suggests that, within our field of view and velocity limits, we have detected all knots with peak surface brightness $S_{peak} > 7\times10^{-16}$ erg cm$^{-2}$ s$^{-1}$ arcsec$^{-2}$. Of the 26 knots at or below this level, 14 are as faint or fainter than nearby features not included in our catalogue.

All of the detected knots are spatially resolved, even though some are quite compact. We estimated the length (maximum dimension) of each knot from the $H_2$ image, and then also the size of the knot perpendicular to its longest dimension to obtain a width/length ratio. These are listed in columns 4 and 5 of Table 2. The FWHM of star images in the $H_2$ mosaic image before continuum subtraction are about 1.0", and a number of the smaller knots are similar in size. However, these knots cannot be stars because they do not appear on the continuum image, and on close examination their profiles on the un-subtracted $H_2$ image have different shapes than the profiles of stars.

The rectangular area shown in Figure 3, and also marked on Figs. 1 and 6, shows the region where our final continuum-subtracted $H_2$ image has its highest (and fairly uniform) signal:noise ratio. We also reduced an additional region from our imaging survey that covers an equally large area to the NE of the portion studied here, but which is mostly the off-Crab sky region and which has considerably lower signal:noise ratio than the region shown in Fig 3b. We could not find any knots in that region. However, on the opposite side of the Crab, on and just off the SW edge of our main region, we do see two additional bright knots at low signal:noise ratio, in an area that is on the Crab Nebula but for which just one or two images contributed to the signal.

## *5. Correlations with other properties*

### *5.1 Comparison of positions of $H_2$ knots to overall Crab morphology*

The $H_2$ knots clearly follow the optical filaments (Fig 6). Even knots 50–52, on the NW side of the nebula (See Fig. 3 for identification), sit on a conspicuous spur of optical emission sticking out from the main system of filaments. This suggests that the knots are physically associated with the filaments, as opposed to being either unrelated condensations in the synchrotron-emitting plasma or part of some sort of amorphous outer shell.

Comparison to Fabry-Perot velocity maps ( Lawrence et al 1995) measuring the [O III] λ5007 emission from these same filaments shows that our $H_2$ velocity coverage is nearly complete except for the high-velocity receding components that lie along the line of sight through the very center of the nebula. This might account for the conspicuous central "hole" in the distribution of $H_2$ knots.

## *5.2 Detailed comparison to HST [S II] and [O III] images*

Further insight about the nature of the molecular knots can be gained by comparing the spatial structure of the $H_2$ emission to that of emission lines from ionized gas, as seen in the HST images. The *I*([S II])/*I*([O III]) intensity ratio is a good indicator of the ionization parameter in the ionized gas if the S/O ratio is constant and the cloud has a $H^+ - H^0$ ionization front, so that both lines are fully formed. Since the knots all receive similar illumination from the synchrotron-emitting plasma, the ionization parameter is mostly a measure of the gas density.

Careful comparison of the morphologies in the three different emission lines shows that most of the $H_2$ knots are associated with regions of high [S II]/[O III] ratio. Figure 7 shows blowups of our $H_2$ image and the HST [S II] and [O III] images with the same scale and position for each knot or group of knots. Column 13 of Table 2 lists our characterization of each $H_2$ knot as either coincident with low-ionization gas ("C"), as immediately adjacent to regions of strong [S II] emission which in turn often have regions of strong [O III] emission on the side opposite from the $H_2$ knot, as would be expected for an ionization front viewed edge-on ("A"), or as filling in the middle of a sheath of [S II] emission ("M"). These categories include 49 of the 52 $H_2$ knots covered by the HST images.

We measured the total [S II] and [O III] signals from the area covered by each $H_2$ knot, using the masks made from the $H_2$ image as described above in § 3. This gave the count rates (in ADU s$^{-1}$) in the [S II] and [O III] lines, the ratio of which we list in column 11 of Table 2. We have not attempted to convert this to a flux ratio both because of the strong velocity-dependence of the correction factors with the rather narrow HST filters, and because the optical spectra described below show that there are multiple emission-line systems along many of these sight-lines. However, as a very rough guide, Figure 8 shows the way in which the inverse of this ratio would depend on velocity for the simple case of narrow lines from a single velocity system where the [O III] λ5007, [S II] λ6717 and [S II] λ6731 lines all have equal strengths. In spite of the uncertainties about measuring an exact [S II]/[O III] ratio from the HST images, it is clear from comparing the two HST images that this ratio is quite high in the immediate vicinities of most of the knots.

It appears that we have only detected the tip of what is likely to be an iceberg of molecular gas permeating the Crab filaments. While looking at very low light levels on an image display to compare the HST images to our continuum-subtracted $H_2$ image, we could see many additional faint $H_2$ features. Some are compact knots similar to the ones we have focused on here, while others are long streamers following the extended linear features seen in the optical emission lines. Many examples of very faint knots may be seen in the panels on Figure 7. The tips of a few of the extended $H_2$ streamers can barely be discerned on Fig. 7 as chains of brighter points in the areas between knots 53 and 54 and below knot 54, between knots 23 and 24, and close to knots 10, 37 and 44; these features are more obvious when seen on an image display.

To evaluate these pervasive, low-level $H_2$–[S II] correlations more quantitatively, we smoothed the $H_2$ and [S II] images into 1.09" pixels and then plotted the pixel-by-pixel intensities against each other. Figure 9a shows the results for all pixels that fall inside the catalogued $H_2$ knots (i.e. for regions inside the boxes shown on Fig. 3c). The large squares and crosses on the figure will be discussed below in § 4.4; it is the small dots representing individual pixel values that are of interest here. The diagonal solid line on Fig. 9a is the lower envelope of [S II]/$H_2$ ratio in the knots, and the more steeply sloping dashed line shows an approximate upper envelope which corresponds to an [S II]/$H_2$ ratio 20 times greater than the lower envelope.

Figure 9b is the same plot, but now showing the pixels in the overlapping parts of the $H_2$ and [S II] images that do *not* fall within the catalogued knots, with the grey scale indicating the logarithm of the number of pixels in each $H_2$, [S II] intensity bin. To eliminate the majority of the pixels that have no $H_2$ emission, we show the difference between the number in a given intensity bin and the number in the bin with the same [SII] intensity and the negative of the $H_2$ intensity. We find that 4 percent of the pixels outside of the catalogued knots show significant $H_2$ emission, and appear to lie within the same range of [S II]/$H_2$ intensity ratios as the pixels inside the catalogued knots. (The cataloged knots cover 2% of the area.) This shows that a significant area of the Crab is covered by $H_2$ emission that correlates with regions that are bright in the [S II] lines.

We measured the total intensity of $H_2$ outside of the cataloged knots from the counts shown in Fig. 9b and the intensity inside the knots from the data shown in Fig 9a, both of which use 1.1-arcsec pixels. The total intensity outside of the cataloged knots is 70% of that inside. The intensity inside the knots is 10% greater than that found in Section 3 by photometering pixels that are 2σ above background when smoothed with a 0.61-arcsec boxcar. The intensities of $H_2$ inside the knots, outside the knots, and combined are $1.5\times10^{-13}$ erg s$^{-1}$ cm$^{-2}$, $1.1\times10^{-13}$ erg s$^{-1}$ cm$^{-2}$, and $2.5\times10^{-13}$ erg s$^{-1}$ cm$^{-2}$, respectively.

### *5.3 Radial velocity measurements*

Since there is a strong correlation between $H_2$ emission and line ratios indicating unusually low ionization, we used the [S II] λ6731 emission line to measure the radial velocities of the $H_2$ knots. We took a grid of long-slit optical spectra with the Goldcam spectrograph on the Kitt Peak National Observatory 2.1m telescope. These spectra will be described in detail in a future paper, but in brief they cover the wavelength range λλ3700-7400Å at 6.3Å resolution with a 3.8" slit width (except for one slit position taken through a 1" slit) and a 5' slit length. For the radial velocity measurements, we extracted a one-dimensional spectrum covering 4.0" along the slit at the expected position of each $H_2$ knot that fell into the slit, after first examining the two-dimensional image on an image display and slightly adjusting the positions of the extraction windows to center them up on any obvious bright knots of strong [S II] emission. At many positions, there were several different velocity systems along the line of sight, in which case we measured the one with the highest *I*([S II] λ6716)/*I*([O III] λ5007) intensity ratio (*i.e.* the lowest-ionization system).

Our optical spectra covered 47 of the $H_2$ knots, for which the measured heliocentric radial velocities are listed in column 12 of Table 2. In almost all of these cases, the [S II] line is stronger than the [O III] line. As a control sample, we extracted spectra at random locations spread all over the Crab. We found *I*([S II])/*I*(O III) ≥ 1 in only 10 of 61 lines of sight which do not contain $H_2$ knots, as compared to *I*([S II])/*I*(O III) ≥ 1 in 41 of the 47 lines of sight that do contain $H_2$ knots. This suggests that in most cases the measured velocities actually do correspond to gas associated with the $H_2$-emitting knots. There was one case (Knot 36) for which there were two velocity systems with large [S II]/[O III] ratios, so Table 2 lists two velocities. Figure 10 shows examples, at the positions of two knots, of the appearance of the [S II] region in the 2D spectrum and the corresponding extracted [S II] and [O III] profiles.

The uncertainties in these radial velocities are significant. For the 13 knots where we had measurements from more than one slit position, the rms scatter in the measured velocities was 32 km s$^{-1}$. However, the instrumental FWHM resolution was 280 km s$^{-1}$ and the exact placement of a knot within the slit leads to velocity uncertainties of about half this value, or ±140 km s$^{-1}$. In addition, examination of the two-dimensional spectral images at the knot locations shows that there are often strong pixel-to-pixel changes in radial velocity, so the effects of different slit positions sampling slightly different places on the sky can be important. An example of this is that here, through a 1" slit, we find a velocity for Knot 1 (the brightest knot) of $v_{helio}$ = +90 km s$^{-1}$, while for Paper I we measured this same knot to have $v_{helio}$ = +161 km s$^{-1}$ using data taken through a 1" slit on a different occasion at a slightly different location on the sky. Both measurements are within the range of velocities seen along the slits for low-ionization gas in the immediate vicinity of the knot. The true velocity uncertainty therefore is roughly ±100 km s$^{-1}$. Our subsequent infrared spectra of seven of these knots (Loh et al. 2011) directly measure the radial velocities of the H$_2$ 2.121 μm line, and they are found to agree well with the [S II] velocities used here, including for one of the blueshifted knots (Knot 46).

Figure 11 shows that the distribution of the radial velocities is very asymmetric. The systemic velocity of the Crab Nebula is very near $v_{helio}$ = 0 (Clarke et al 1983 found $v$ = -20 km s$^{-1}$, with an uncertainty which we estimate to be ±50 km s$^{-1}$). In spite of the velocity uncertainties described above, it is clear that most of the knots that we can see are located on the receding (far) side of the nebula. This shows that the H$_2$ emission is strongest on the side of the clouds closest to the pulsar, and that the dust extinction across the clouds is large at 2 μm. We will return to these points in a future paper.

The few knots with negative radial velocities do not have any other characteristics that are obviously different from the knots with positive radial velocities. This is shown on Figure 12, where the mean and peak surface brightness S(H$_2$) and the knot length from Table 2 are plotted as a function of the knot's radial velocity.

Therefore there is no evidence of an instrumental bias against knots with negative velocities. The knots with positive and negative velocities do not segregate in the plot of the intensities of H$_2$ and [S II] (Fig 9a). Four of nine of the knots with negative velocity lie near the line showing knots with the brightest H$_2$/[S II] ratio.

### *5.4 The Brγ image*

The Brγ emission line is much fainter than the H$_2$ knots and was right at our detection limit. The emission that we were able to measure from this H recombination line shows up as a much more amorphous component than the H$_2$ knots. In Fig. 3b, the locations of the H$_2$ knots are shown as rectangles on top of the Brγ image. It is clear that even though a number of the H$_2$ knots lie along or next to Brγ -emitting filaments, the Brγ emission does not have the clumped structure of the H$_2$ knots. Rather, it primarily traces some of the very extended filaments that are visible in Balmer line and [O III] images (compare to Fig. 6), as would be expected.

As was described above, the filter cutoff at large wavelengths has a strong effect on the Brγ line. Comparing to places where [O III] emission is actually seen at specific velocities on the Lawrence et al. (1995) [O III] Fabry-Perot maps, we find that the velocity selection effects cause our Brγ image to systematically miss the far side of the Crab's "He-rich ring" (Uomoto & MacAlpine 1987; Lawrence et al. 1995) where it extends to the SW of the pulsar. The Brγ map also is not sensitive to some of the bright filaments running perpendicular to that ring to the north of it, nor to the highest-velocity receding gas on lines of sight near the pulsar. However, in terms of the actual $H_2$ knots in our catalog, using the velocities determined above, only knots 19, 24 and perhaps 44 fall outside the Brγ velocity range.

There is a correlation between the $F(Brγ)/F(H_2)$ values for the areas covered by the knots (Table 2, column 10), and the [S II]/$H_2$ ratios shown in Fig. 9a. In that figure, the points corresponding to peak intensities for knots with $F(Brγ)/F(H_2) > 0.5$ are surrounded by squares, while those for knots with $F(Brγ)/F(H_2) < 0.25$ are marked with large "+" symbols. Clearly, the knots with high (or low) $F(Brγ)/F(H_2)$ also have high (or low) [S II]/$H_2$. This is somewhat surprising, since [S II] and $H_2$ are spatially correlated, but Brγ and $H_2$ are not. However, this correlation in intensity is likely to be dominated by the fact that [S II] comes from gas that also emits Brγ, so that both the Brγ and [S II] emission are indicating the same connection between the molecular $H_2$-emitting gas and adjacent ionized gas.

The main result from the Brγ image is from the morphology. It sets a rather stringent upper limit on the $I(Brγ)/I(H_2\ \lambda 2.12\mu m)$ intensity ratio from the actual $H_2$ knots. Even though column 10 of Table 2 shows detections rather than limits for Brγ at the positions of a number of knots, this still only places an upper limit on the Brγ/$H_2$ intensity ratio that should be used to study the structure and physical conditions in individual molecular knots, because the morphology clearly shows that the measured Brγ is not coming from the dense knots.

## *6. Conclusions*

Our near-IR survey of the Crab Nebula has found 55 molecular knots that emit strongly in the $H_2$ 2.12μm line. The knots generally fall on or next to optically-bright filaments. Comparison of the positions and structures of the $H_2$ knots to the morphology seen in HST optical-passband [S II] and [O III] images shows that most $H_2$ knots are associated with low-ionization (large $I([S\ II])/I([O\ III])$) gas (although the reverse is not necessarily true; so far $H_2$ emission has been detected from only a small fraction of the low-ionization regions in the Crab). There are also a few cases where $H_2$ emission is seen coming from gas that does not emit strongly in [S II].

Guided by the tendency for $H_2$ emission to come from regions of strong [S II] emission, we measured the velocity of the [S II] emission lines at the locations of 47 of the 55 $H_2$ knots. In 75 percent of the cases, the [S II] line has a positive radial velocity relative to the Crab's systemic velocity, so the $H_2$ knots are preferentially seen on the far side of the nebula.

These knots are just the brightest examples from a distribution that fades down into the background noise in our NIR images. Close comparison with the HST images suggests that there is also $H_2$ emission from faint filamentary structures following the optically-bright filaments. A pixel-by-pixel analysis shows that about 6 percent of the Crab's projected surface area has significant $H_2$ emission that correlates with [S II] emission.

The presence of $H_2$ indicates the presence of [S II], but the ratio of $I([S\ II])/I(H_2)$ varies, by up to possibly a factor of 20. Furthermore, the ratio is correlated with the ratio of $I(Brγ)/I(H_2)$

We also detected weak Brγ emission, but it clearly comes from the general distribution of ionized gas all along the filaments rather than from the molecular gas that produces the bright knots of $H_2$ emission. The limited bandpass of our Brγ filter cuts off some velocities from the far side of the nebula, but this only affects 2 or 3 of the bright $H_2$ knots that we describe here. The Brγ/$H_2$ intensity ratio is quite low for the remaining knots, with $I(Brγ)/I(H_2) < 1$ for 50 of the remaining 52 cases (as compared to $I(Brγ)/I(H_2) \sim 10$ for a typical H II region and PDR such as in the Orion Nebula).

It is not a great surprise to find $H_2$ in a large number of Crab filaments. G90 found $H_2$ emission from two of the three locations that they studied, and it has often been suggested that the dust blobs seen in absorption against the Crab's synchrotron continuum should also contain $H_2$ (Fesen & Blair 1990, Sankrit et al 1998 ). The strong correlation between $H_2$ knots and bright concentrations of [S II] is also expected, since high-ionization lines associated with the Crab's filaments often come from diffuse regions surrounding more concentrated zones emitting low-ionization lines ([S II], [O I]) and having central dusty cores (Blair et al. 1997; Sankrit et al 1998). However, our work is the first study at relatively high angular resolution of the actual location and structure of the molecular gas.

We now have enough information about the $H_2$ distribution so that we can begin a careful comparison to the distribution and morphology of the absorbing dust blobs. The $H_2$ has most likely been formed by catalysis on dust, so the two should be intimately mixed together. We find an asymmetrical distribution of radial velocities, in which most of the $H_2$ emission comes from knots on the far side of the Crab's expanding SN shell. This implies that most of the similar number of molecular knots which ought to exist on the near side are hidden from view, with internal dust absorption being the likely mechanism. We are currently in the process of comparing the dust absorption and molecular emission properties of the Crab to test this idea and to learn more about the geometry of the individual knots.

Separate papers will also discuss the ramifications of these results for understanding the physical nature of the molecular knots. Possible excitation mechanisms have already been discussed by G90 and Loh et al. (2010). Our optical spectroscopy will be combined with K-band infrared spectra to test the relative importance of UV fluorescence, shocks, and heating by fast electrons, with the aid of detailed models using our *Cloudy* plasma simulation code (last reviewed by Ferland et al. 1998).

A fuller understanding of the basic means of excitation and of the densities and temperatures in these molecular cores will provide insight into important details of this particular core-collapse supernova explosion. These include the mass budget of the ejecta; Loh et al (2010) showed that there may be considerable mass hiding in cool, hard-to-detect molecular gas in addition to the 2–5$M_\odot$ that is conventionally measured mostly from the ionized gas (Fesen, Shull & Hurford 1997). Spectroscopy of the molecular gas can also address the dynamical evolution of the filaments and hence of the ejected material in general.

Studies of these dense cores also bear on a series of bigger-picture questions. They permit an improved inventory of the chemical abundances (and abundance variations) in this gas that is being returned to the ISM. The filaments contain dust, and can show us the details of how supernovae produced the dust that obscures the first epoch of star formation (see Silk & Rees 1998; Michalowski et al. 2010). Finally, the Crab Nebula filaments may be nearby, small-scale examples of the filamentary systems observed in cool-core clusters of galaxies. There are many properties in common, most notably the presence of warm molecular emission in filaments surrounded by hostile ionizing particles (Fabian et al. 2008; Salomé et al. 2008; Ferland et al. 2009).

*Acknowledgements.* EDL, JAB and ZKC are grateful for support from NASA through ADP grant NNX10AC93G. GJF acknowledges support from NSF (0908877), NASA (07-ATFP07-0124 and 10-ATP10-0053) and STScI (HST-AR-12125.01).## *References*

Bietenholz, M. F., Kronberg, P. P., Hogg, D. E., Wilson, A. S. 1991. Ap J, 373, L59

Blair, W.P., Davidson, K., Fesen, R.A., Uomoto, A., MacAlpine, G.M. & Henry, R.C.B. 1997, ApJS, 109, 473.

Campins, H., Rieke, G. & Lebofsky, M., 1984, AJ, 90, 896.

Clark, D., Murdin, P., Wood, R., Gilmozzi, R., Danziger, J., & Furr, A. 1983, MNRAS, 204, 415.

Covey, K. R., 2007, AJ, 134, 2398.

Davidson, K. & Fesen, R.A. 1985, ARA&A, 23, 119

Davidson, K. 1973, ApJ, 186, 223

Davidson, K. 1978, ApJ, 220,177

Fabian, A.C. et al. 2008, Nature, 454, 968

Ferland, G.J., Fabian, A.C., Hatch, N.A., Johnstone, R.M., Porter, R.L., van Hoof, P. A. M. & Williams 2009, R.J.R., MNRAS, 392, 1475

Ferland, G. J.; Korista, K. T.; Verner, D. A.; Ferguson, J. W.; Kingdon, J. B.; Verner, E. M. 1998, PASP, 110, 761

Fesen, R. A. & Blair, W.P. 1990, ApJ, 351, L45

Fesen, R.A., Shull, J.M. & Hurford, A.P. 1997, AJ, 113, 354

Graham, J.R., Wright, G.S. & Longmore, A.J. 1990, ApJ, 352, 172 (G90)

Henry, R. B. C., & MacAlpine, G. M. 1982, ApJ, 258, 11

Hester, J.J. 2008, ARA&A, 46, 127

Kurucz, R., 1979, ApJS, 40, 1.

Labs, D., & Neckel, H., 1968, Z. Astrophysik, 69, 1.

Lawrence, S.S., MacAlpine, G.M., Uomoto, A. et al 1995, AJ, 109, 2635

Loh, E. D., & Abdul Rashid, A. A., 2010, http://www.pa.msu.edu/~loh/SpartanIRCamera/Astrometry.pdf

Loh, E.D., Baldwin, J.A. & Ferland, G.J. 2010, ApJ, 716, L914

*Figures*

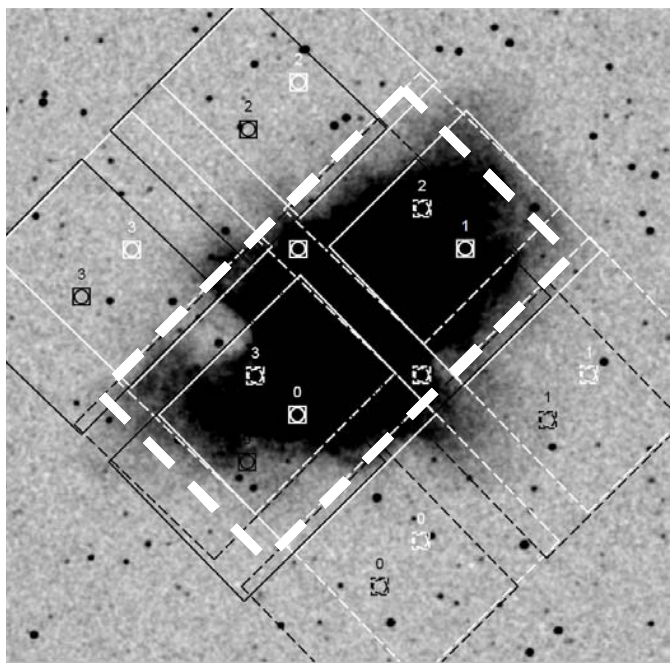

Figure 1. Detector patterns for targets A and A' (dashed black and white lines) and B and B' (solid black and white lines) superposed on the 2MASS K-band image. The area shown is 7.8' wide, with North up and East to the left. The very heavy white dashed rectangle indicates the area used in our final analysis (as shown below in Figs 3, 4 and 6).

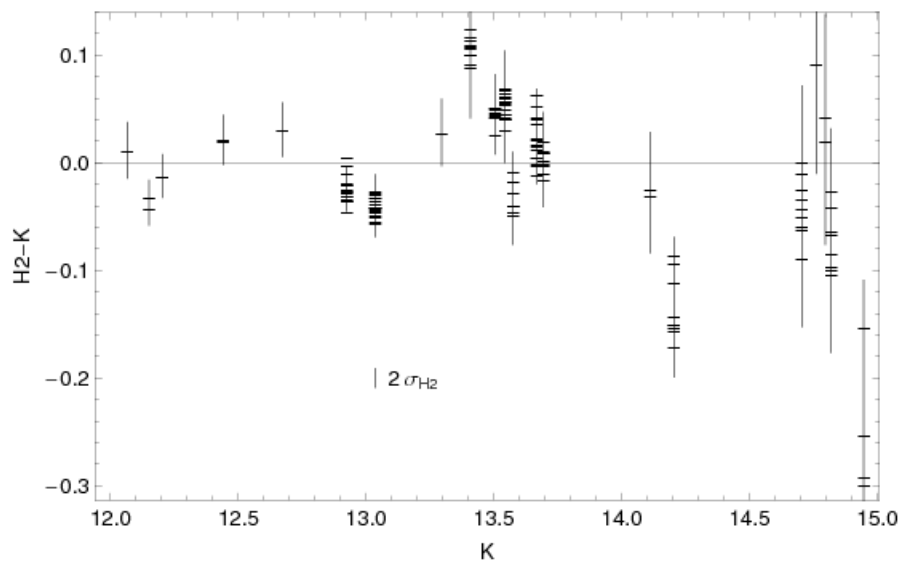

Figure 2. Difference between the H2 and 2MASS K-band photometry of the stars used for calibration. The error bars account only for the error in the 2MASS photometry. Also shown is the error of a single H2 measurement of a 13th mag star.

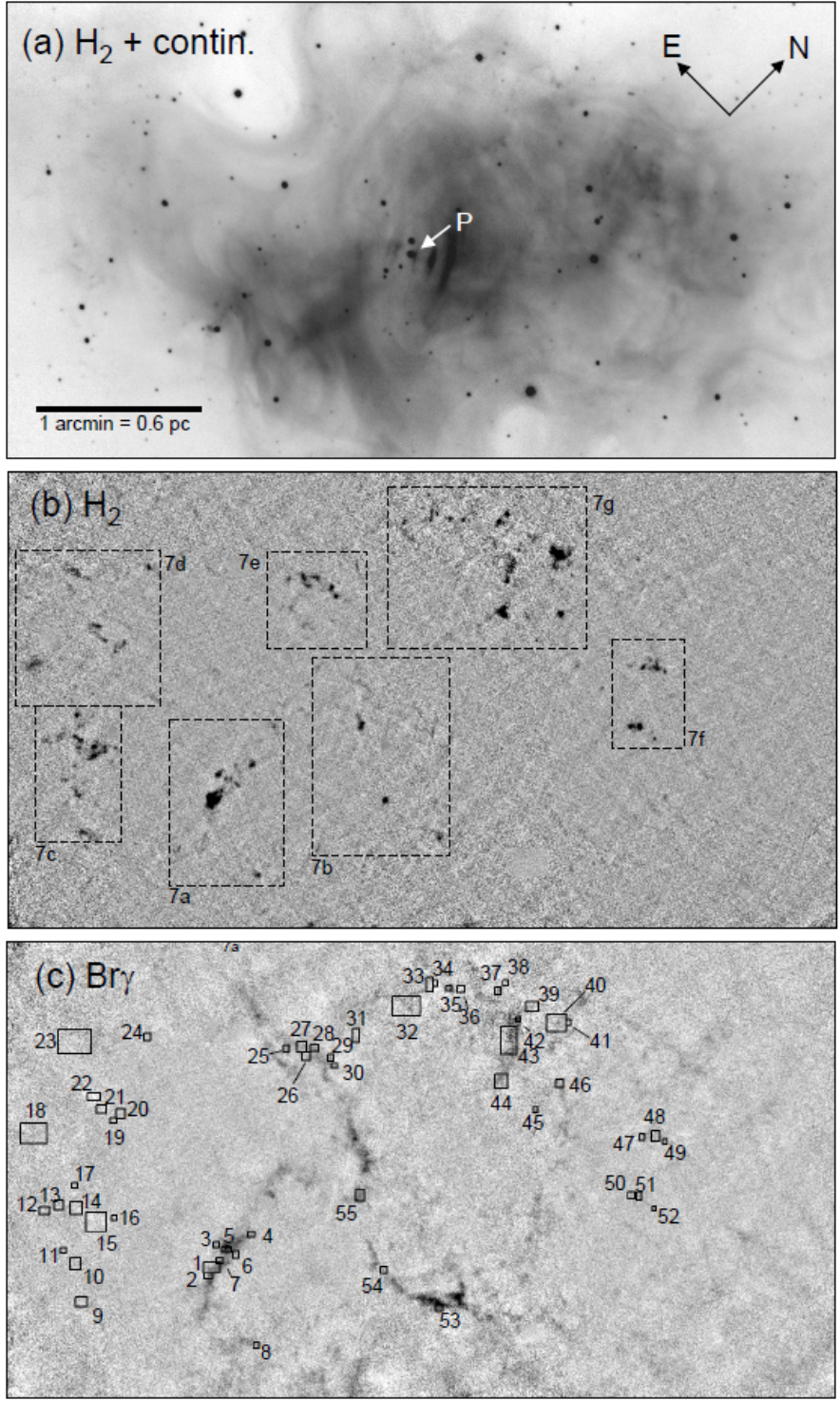

Figure 3. Spartan camera images in the $H_2$ 2.12 μm filter before (panel a) and after (panel b) continuum subtraction, and the continuum-subtracted Brγ image (panel c). The area on the sky covered by these three images is shown on Figures 1 and 6. The dashed rectangles in panel b show the areas covered in panels a–g of Fig. 7, while the numbered rectangles on the Brγ image show the areas in our measurements of each individual $H_2$ knot.

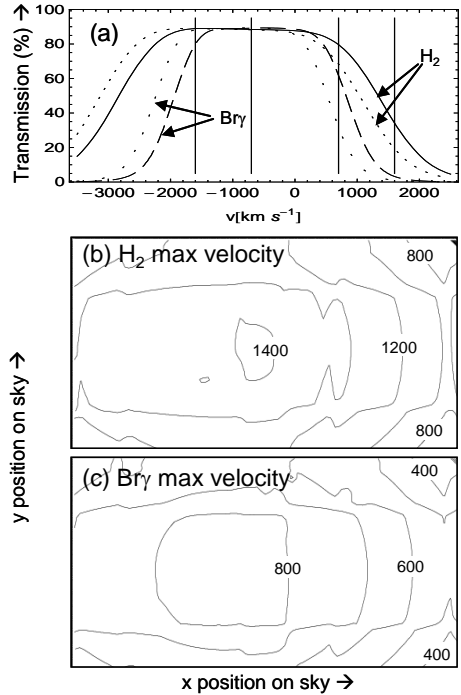

Figure 4. *(a)* Spartan Camera filter transmission curves. The heavier curves are for on-axis transmission, while the dotted curves are for the edge of the field. The vertical lines show the range of Crab expansion velocities, as described in the text. *(b)* Contours of the high-velocity cutoff (in km s$^{-1}$) for $H_2$. The rectangle covers the same area on the sky as each image shown in Fig. 3. *(c)* Contours of the high-velocity cutoff for Brγ.

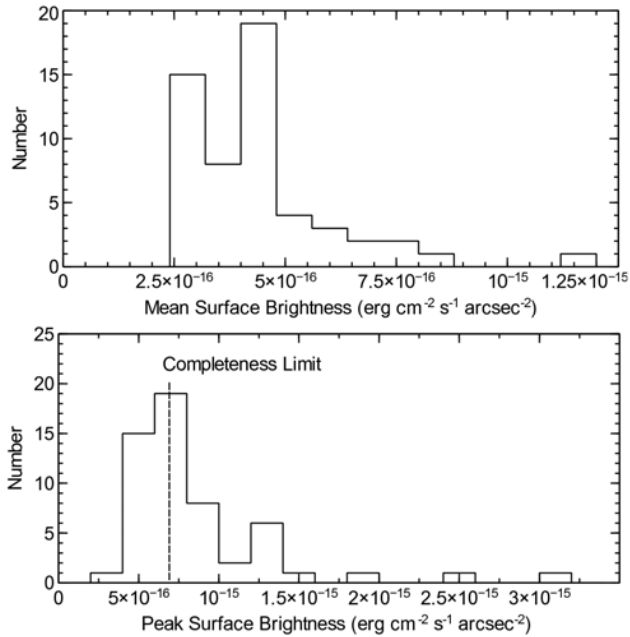

Figure 5. Histograms of mean and peak surface brightness of $H_2$ knots.

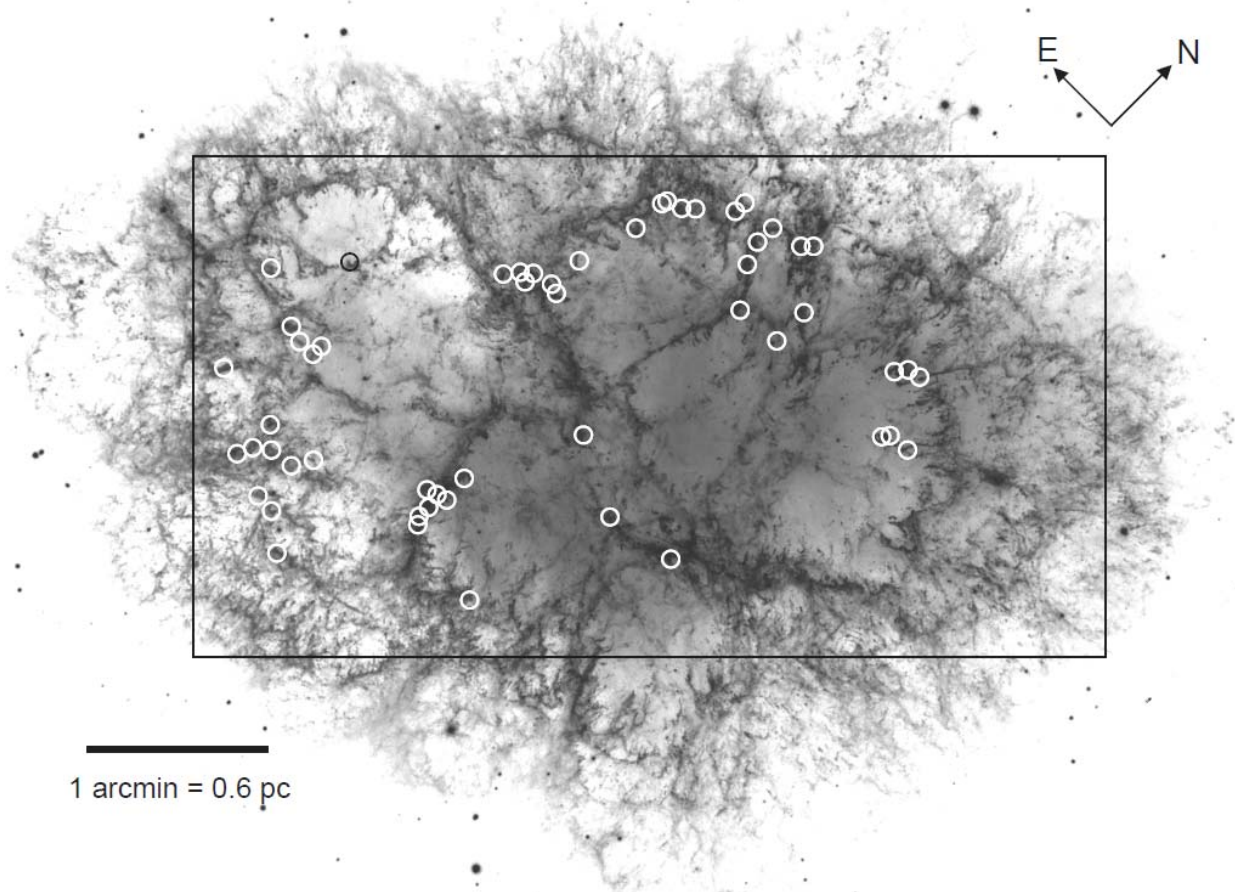

Figure 6. Positions of H$_2$ knots shown as circles. The background image, which shows the full Crab, is a grey-scale rendering of the well-known HST composite color image made with F502N, F631N and F673N filters (courtesy of NASA, ESA, J. Hester, and A. Loll (Arizona State University); STScI News Release 2005-37). The rectangle shows the area covered by the Spartan Camera images in Fig. 3.

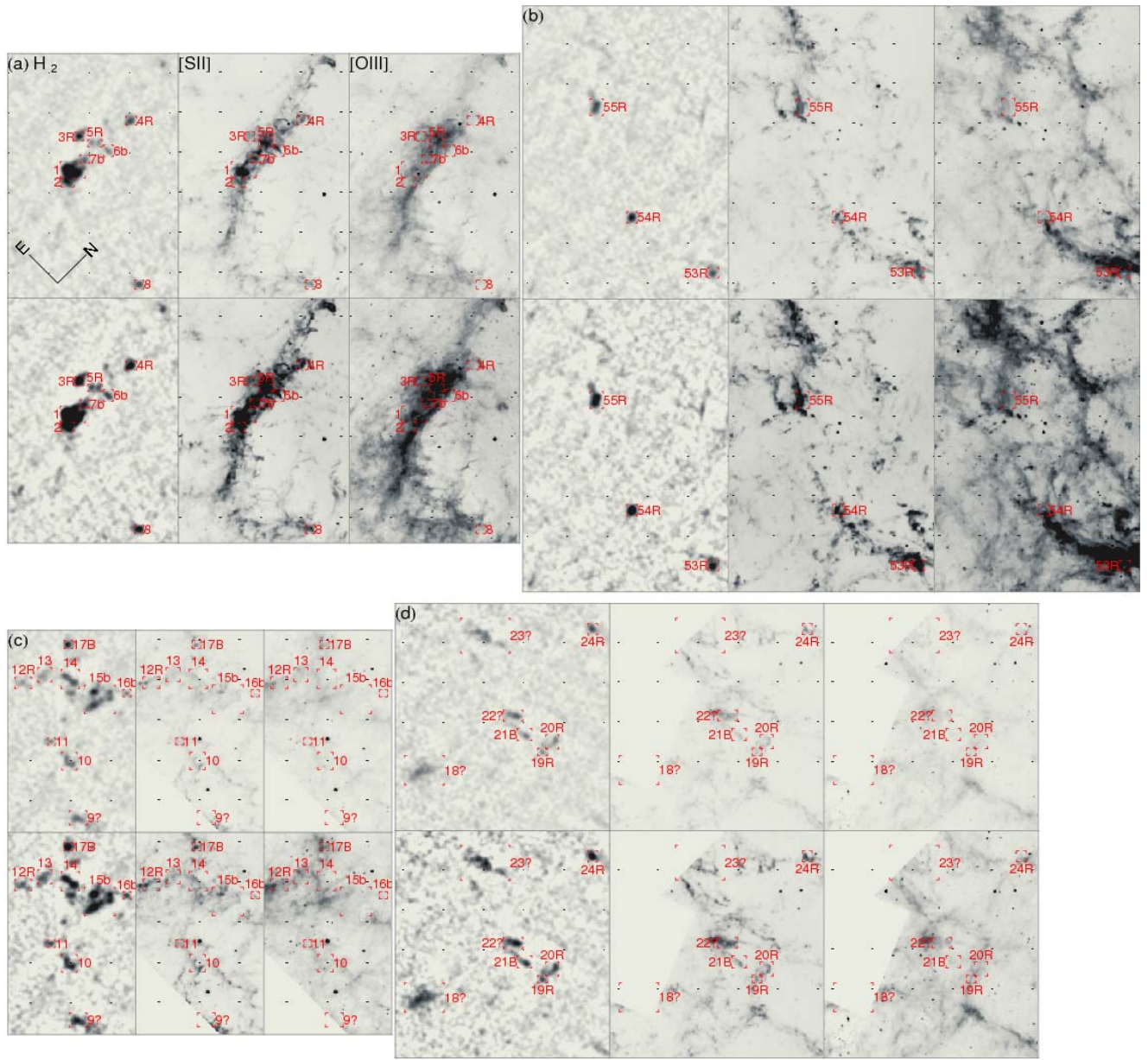

Figure 7a. Each panel (a-d) shows six blowups of the same set of knots, as seen in three different emission lines and at two different contrast levels. The area covered on the sky by each of these panels is indicated on Fig. 3b. The emission lines, $H_2$ [S II] and [OIII], are indicated at the top of panel (a). For comparison between panels, the images in each emission line in the upper row of each panel are all shown with the same intensity range, and those in the frame below cover the faint half of that intensity range. All panels have the same spatial scale. The grid spacing as well as the lengths of the compass legs is 10". Panel (a) shows knots 1–8, panel (b) knots 53–55, panel (c) knots 9–17, and panel (d) knots 18–24. The symbols following the knot's numbers indicate the radial velocity: "B" for $v < 300$ km s$^{-1}$; "b" for $-300 - <v<0$ km s$^{-1}$; no symbol for $0 < v < +300$ km s$^{-1}$; "R" for $v > 300$ km s$^{-1}$; "m" for multiple velocities; "?" for no velocity information.

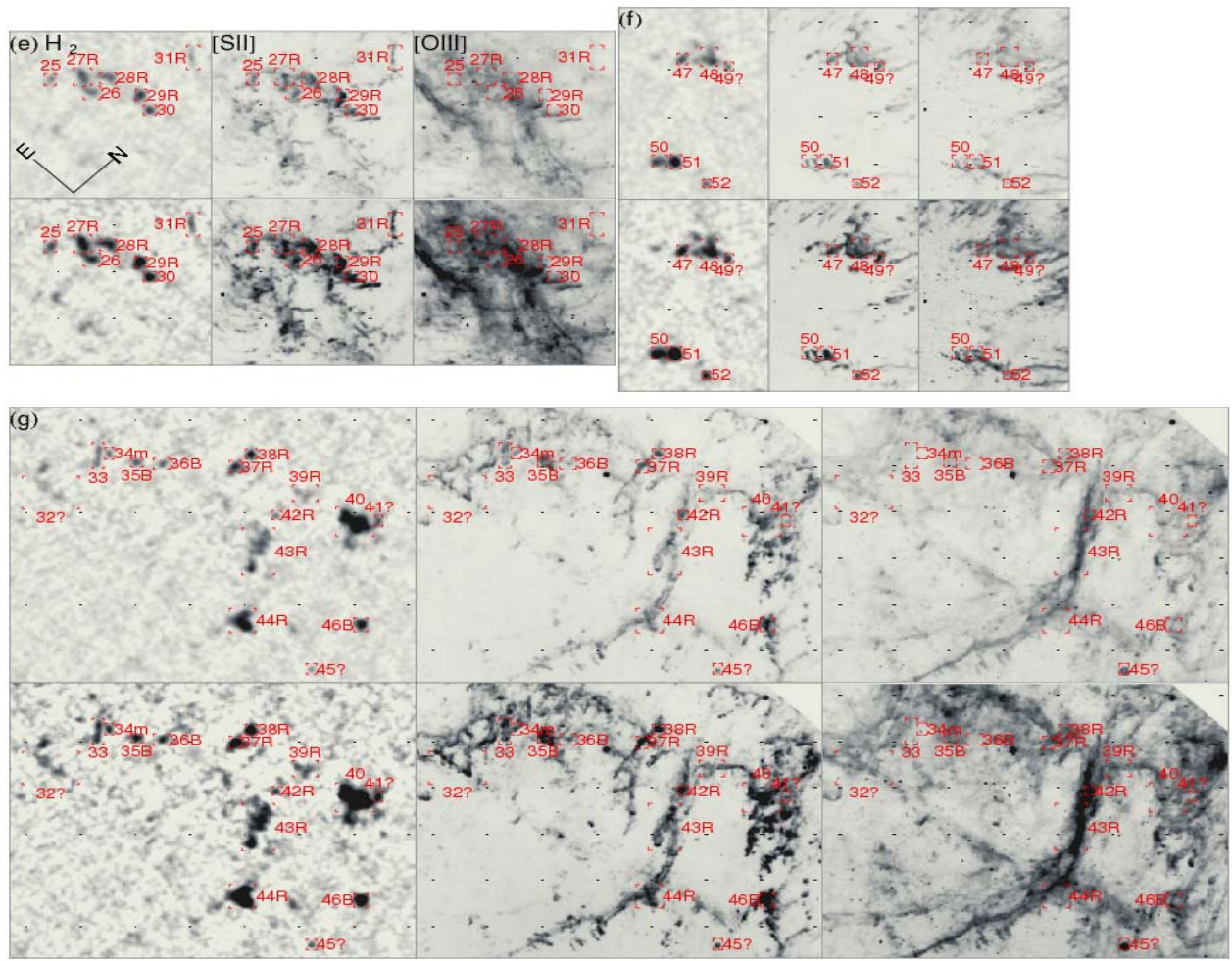

Figure 7b. Continuation of Fig. 7a, with the same spatial and intensity scales. Panel (e) shows knots 25–31, panel (f) knots 47–52, and panel (g) knots 32–46.

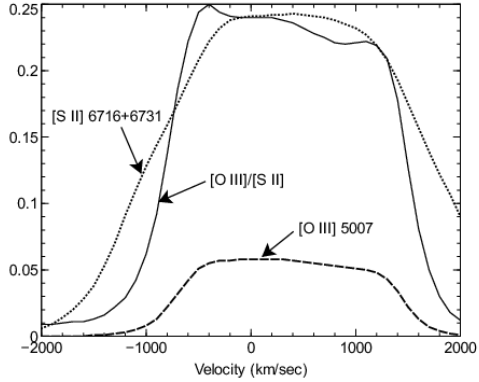

Figure 8. Factor to convert ratios of [S II]/[O III] count rates into flux ratios (solid line). The dashed and dotted lines show the HST throughput for the [OIII] $\lambda 5007$ and [S II] $\lambda 6716+\lambda 6731$ lines, respectively,, from the STSDAS *Synphot* package in IRAF.

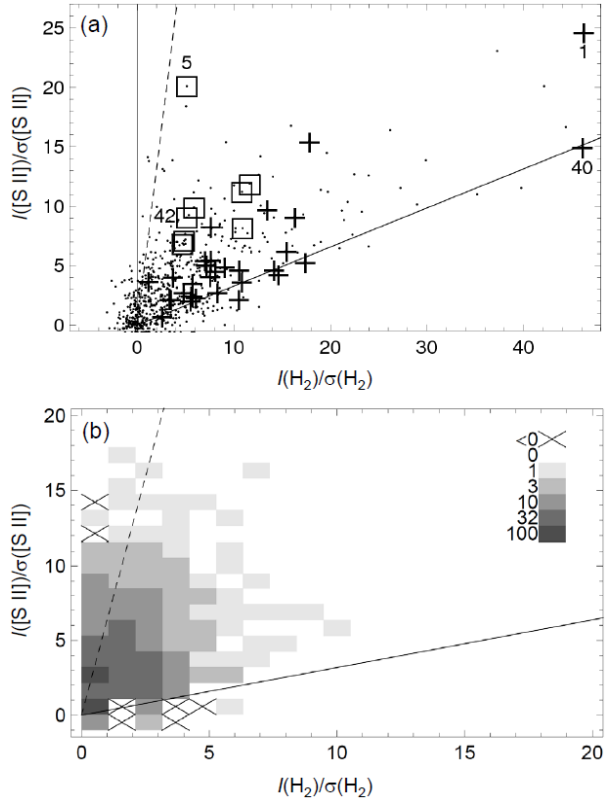

Figure 9. *(Panel a)* The small dots indicate pixel intensity of $H_2$ and [S II] in the regions that fall within the catalogued $H_2$ knots, in units of the standard deviation of the pixel values over the entire image. The solid line goes through the brightest pixel in $H_2$ for knot 40, and the ratio of [S II] to $H_2$ is 20 times greater for the dashed line. A few other knots are labeled. The squares indicate knots with $F(Br\gamma)/F(H_2) > 0.6$, and the "+" symbols those with $F(Br\gamma)/F(H_2) < 0.25$. *(Panel b)* Same as *(a)*, but now showing the pixels in the overlapping parts of the $H_2$ and [S II] images that do *not* fall within the catalogued knots. The grey scale is proportional to the logarithm of the number of pixels in each $H_2$, [S II] intensity bin.

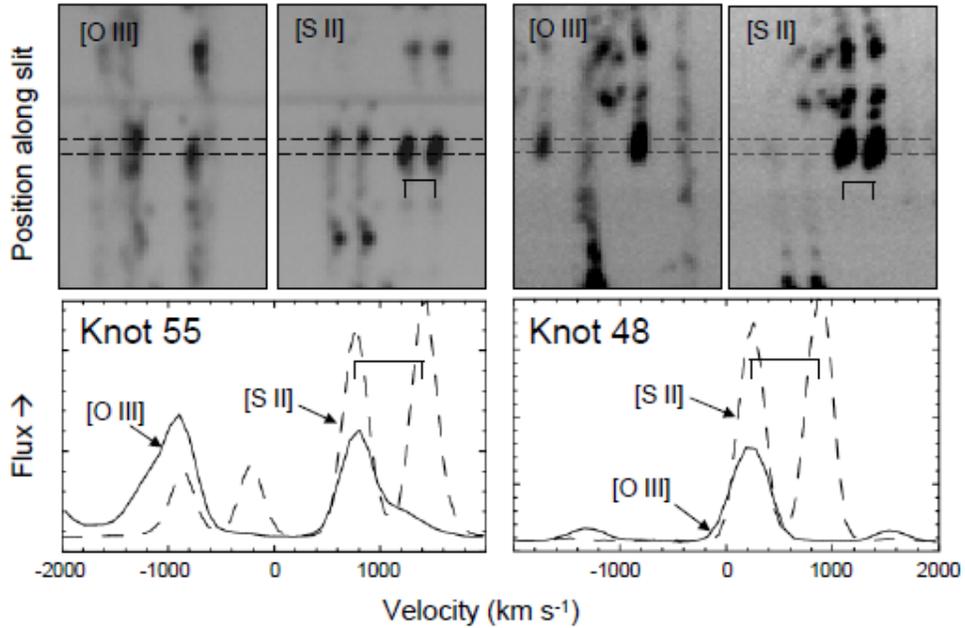

Figure 10. Examples, at the positions of knots 55 and 48, of the appearance of the [O III] and [S II] regions in the 2D optical spectra, and the corresponding extracted [O III] (solid line) and [S II] (dashed line) profiles. The dashed horizontal lines in the 2D images mark the slit position, and the double tick marks on the [S II] images and extracted spectra show the separation of the [S II] $\lambda\lambda$6716, 6731 doublet. The velocity scale is for the [O III] $\lambda$5007 and [S II] $\lambda$6716 lines.

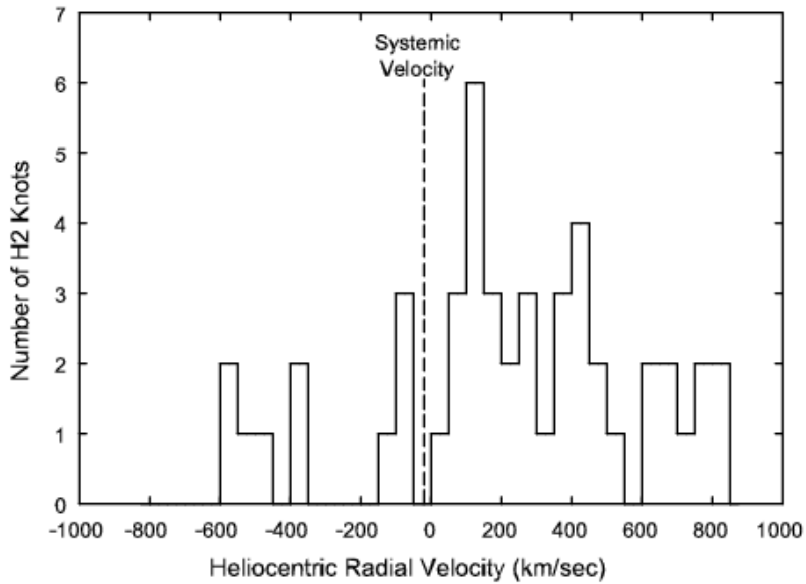

Figure 11. Histogram of radial velocities of the $H_2$ knots.

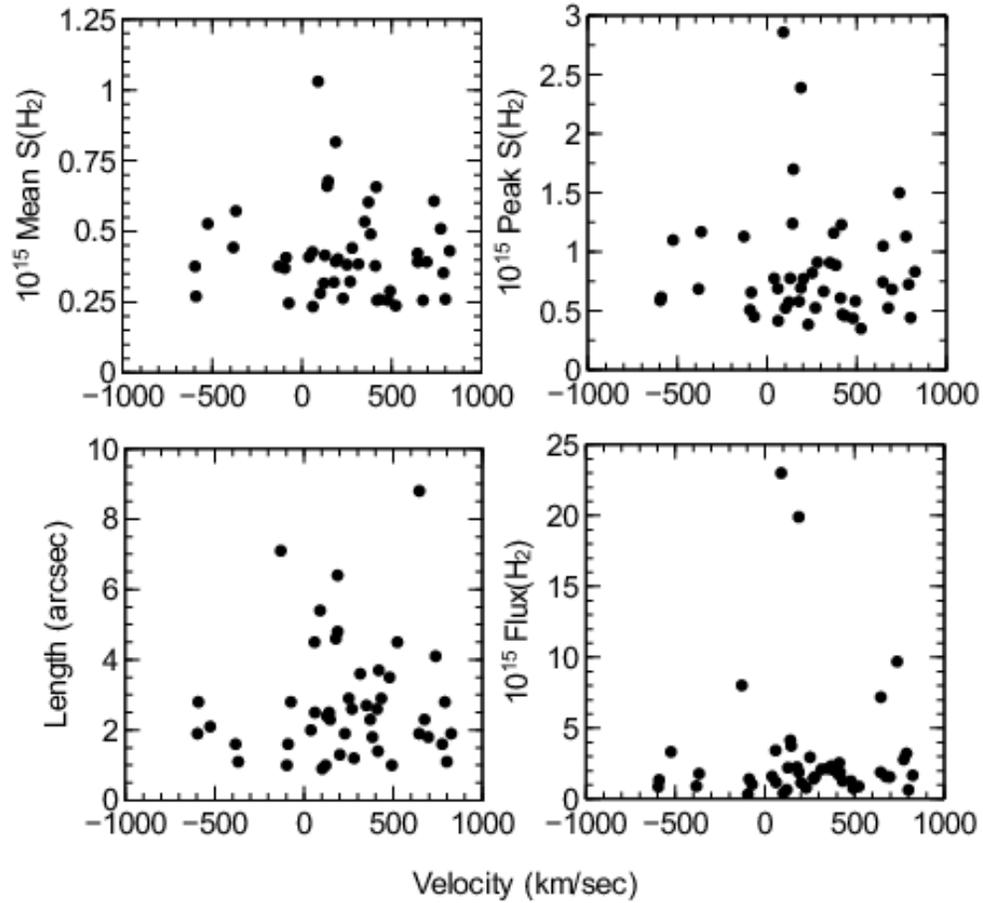

Figure 12. Knot parameters as a function of radial velocity. The mean and peak surface brightness (top-left and top-right, respectively) are in units $10^{15}$ erg cm$^{-2}$ s$^{-1}$ arcsec$^{-2}$. The total flux from each knot (bottom-right) is in units $10^{15}$ erg cm$^{-2}$ s$^{-1}$. A knot's length is its longest dimension as projected on the sky.

*Tables*

| Table 1. Observing log Spartan Infrared Camera on the SOAR Telescope | | | | |
|---|---|---|---|---|
| Date (UT) | No. of $H_2$ images | No. of Cont3 images | No. of Br$\gamma$ images | Observing conditions |
| 2009 Dec 24 | 24 | 32 | 30 | 0.7" FWHM images, high humidity |
| 2009 Dec 25 | 39 | 40 | 40 | 0.7"FWHM images, photometric |
| 2010 Jan 04 | 39 | 39 | 36 | 1.3" FWHM images |
| 2010 Feb 07 | 12 | 16 | 16 | 2" elongated images, high wind |
| 2010 Mar 24 | 15 | 12 | 12 | 1.5" FWHM images. |
| Total images | 129 | 139 | 134 | |
| Total exp. time | 6.45 hr | 6.95 hr | 9.31 hr | |

Table 2. Catalogue of Bright $H_2$ Knots in the Crab Nebula

| Knot | RA J2000 | Dec J2000 | Length (arcsec) | W/L | Area (arcsec$^2$) | $S_{avg}(H_2)$ (erg cm$^{-2}$ s$^{-1}$ arcsec$^2$) | $S_{peak}(H_2)$ (erg cm$^{-2}$ s$^{-1}$ arcsec$^2$) | $F(H_2)$ (erg cm$^{-2}$ s$^{-1}$) | $F(Br\gamma)/F(H_2)$ | [S II]/[O III] count rates | $v_{helio}$ (km s$^{-1}$) | Morph | Notes |
|---|---|---|---|---|---|---|---|---|---|---|---|---|---|
| 1 | 05:34:34.31 | 21:59:40.69 | 5.4 | 0.4 | 22.4 | 1.13E-15 | 3.14E-15 | 2.53E-14 | 0.16 | 13 | 90 | C | |
| 2 | 05:34:34.22 | 21:59:38.17 | 2.5 | 0.4 | 6.3 | 7.21E-16 | 1.35E-15 | 4.53E-15 | 0.31 | 12 | 141 | C | |
| 3 | 05:34:34.65 | 21:59:48.17 | 1.4 | 0.8 | 3.9 | 6.94E-16 | 1.29E-15 | 2.69E-15 | 0.13 | 8 | 414 | C | |
| 4 | 05:34:34.20 | 21:59:59.89 | 1.9 | 0.6 | 4.5 | 4.36E-16 | 7.69E-16 | 1.95E-15 | 0.25 | 18 | 645 | C | |
| 5 | 05:34:34.37 | 21:59:49.73 | 2.9 | 0.5 | 4.9 | 2.73E-16 | 4.84E-16 | 1.33E-15 | 1.40 | 11 | 433 | | |
| 6 | 05:34:34.10 | 21:59:50.69 | 2.8 | 0.4 | 4.2 | 2.77E-16 | 5.10E-16 | 1.16E-15 | 0.39 | 13 | -73 | C | |
| 7 | 05:34:34.29 | 21:59:44.95 | 1.6 | 0.8 | 3.5 | 4.59E-16 | 7.42E-16 | 1.60E-15 | 0.44 | 12 | -89 | C | |
| 8 | 05:34:32.01 | 21:59:32.45 | 2.0 | 1.0 | 3.9 | 4.47E-16 | 8.47E-16 | 1.75E-15 | 0.17 | 17 | 39 | C | multiple spatial components |
| 9 | 05:34:36.11 | 21:58:57.77 | 3.0 | 0.4 | 5.7 | 3.86E-16 | 6.86E-16 | 2.21E-15 | <0.14 | 17 | | - | |
| 10 | 05:34:36.94 | 21:59:06.31 | 4.6 | 0.5 | 7.0 | 3.46E-16 | 6.29E-16 | 2.43E-15 | <0.12 | 15 | 178 | C | |
| 11 | 05:34:37.43 | 21:59:06.71 | 0.9 | 1.0 | 1.5 | 3.07E-16 | 5.72E-16 | 4.51E-16 | <0.61 | 14 | 102 | C | |
| 12 | 05:34:38.51 | 21:59:11.87 | 3.5 | 0.5 | 5.0 | 2.67E-16 | 4.57E-16 | 1.33E-15 | <0.22 | 7 | 479 | | $H_2$ is at cusp where two velocity systems join |
| 13 | 05:34:38.36 | 21:59:17.32 | 2.5 | 1.0 | 5.0 | 2.57E-16 | 4.59E-16 | 1.30E-15 | <0.22 | 7 | 61 | C | |
| 14 | 05:34:37.96 | 21:59:20.92 | 4.5 | 0.2 | 8.0 | 4.70E-16 | 7.59E-16 | 3.77E-15 | <0.08 | 10 | 59 | C | |
| 15 | 05:34:37.38 | 21:59:22.33 | 7.1 | 0.6 | 21.3 | 4.25E-16 | 1.28E-15 | 9.07E-15 | <0.03 | 11 | -130 | C | |
| 16 | 05:34:37.09 | 21:59:28.46 | 1.0 | 1.0 | 0.9 | 4.17E-16 | 5.73E-16 | 3.65E-16 | <0.64 | 13 | -96 | C | multiple spatial components |
| 17 | 05:34:38.44 | 21:59:26.67 | 1.1 | 1.0 | 3.1 | 6.64E-16 | 1.36E-15 | 2.08E-15 | <0.13 | 8 | -368 | A | |
| 18 | 05:34:40.20 | 21:59:28.96 | 6.3 | 0.3 | 15.4 | 2.75E-16 | 5.32E-16 | 4.22E-15 | 0.24 | 7 | | - | |
| 19 | 05:34:38.91 | 21:59:53.63 | 1.1 | 1.0 | 2.5 | 2.65E-16 | 4.52E-16 | 6.60E-16 | <0.35 | 10 | 800 | A | |
| 20 | 05:34:38.90 | 21:59:57.28 | 2.3 | 0.4 | 6.2 | 2.63E-16 | 5.40E-16 | 1.63E-15 | <0.13 | 10 | 675 | C | |
| 21 | 05:34:39.35 | 21:59:53.54 | 2.8 | 0.5 | 5.0 | 3.19E-16 | 7.27E-16 | 1.59E-15 | <0.16 | 14 | -591 | C | |
| 22 | 05:34:39.75 | 21:59:54.90 | 3.8 | 0.3 | 7.8 | 3.76E-16 | 7.63E-16 | 2.93E-15 | -0.19 | 16 | | C | |
| 23 | 05:34:41.16 | 22:00:03.81 | 9.8 | 0.3 | 28.0 | 2.62E-16 | 6.57E-16 | 7.35E-15 | 0.14 | 13 | | - | |
| 24 | 05:34:39.86 | 22:00:24.28 | 1.9 | 0.6 | 3.9 | 4.39E-16 | 8.47E-16 | 1.70E-15 | <0.13 | 19 | 824 | C | |

| | | | | | | | | | | | | |
|---|---|---|---|---|---|---|---|---|---|---|---|---|
| 25 | 05:34:37.01 | 22:00:57.80 | 1.9 | 0.7 | 3.0 | 2.85E-16 | 4.17E-16 | 8.55E-16 | 0.59 | 12 | 229 | A | |
| 26 | 05:34:36.51 | 22:01:01.21 | 2.6 | 0.5 | 4.4 | 3.47E-16 | 5.69E-16 | 1.53E-15 | 0.24 | 10 | 269 | | $H_2$ sits in [S II], [O III] hole. |
| 27 | 05:34:36.75 | 22:01:02.41 | 3.6 | 0.4 | 5.5 | 4.11E-16 | 7.14E-16 | 2.25E-15 | 0.24 | 9 | 315 | A | |
| 28 | 05:34:36.51 | 22:01:05.62 | 2.6 | 0.5 | 4.6 | 4.00E-16 | 6.45E-16 | 1.86E-15 | 0.32 | 10 | 409 | A | |
| 29 | 05:34:36.02 | 22:01:06.91 | 1.8 | 0.7 | 4.2 | 5.20E-16 | 9.42E-16 | 2.18E-15 | 0.27 | 20 | 383 | C | |
| 30 | 05:34:35.77 | 22:01:05.90 | 1.2 | 1.0 | 3.6 | 4.75E-16 | 9.81E-16 | 1.69E-15 | 0.19 | 16 | 280 | A | |
| 31 | 05:34:35.97 | 22:01:19.45 | 4.5 | 0.3 | 3.8 | 2.48E-16 | 3.66E-16 | 9.31E-16 | <0.27 | 21 | 523 | A | |
| 32 | 05:34:35.55 | 22:01:40.73 | 6.6 | 0.2 | 13.4 | 2.42E-16 | 5.40E-16 | 3.25E-15 | 0.61 | 13 | | A | |
| 33 | 05:34:35.50 | 22:01:51.99 | 4.8 | 0.2 | 4.8 | 4.21E-16 | 7.46E-16 | 2.02E-15 | <0.17 | 23 | 188 | M | |
| 34 | 05:34:35.43 | 22:01:53.88 | 2.2 | 0.4 | 1.9 | 4.41E-16 | 6.89E-16 | 8.40E-16 | <0.42 | 31 | -474,101 | A | |
| 35 | 05:34:35.08 | 22:01:56.12 | 1.6 | 1.0 | 2.1 | 5.12E-16 | 7.92E-16 | 1.05E-15 | 0.62 | 17 | -383 | C | |
| 36 | 05:34:34.81 | 22:01:59.33 | 1.9 | 0.7 | 2.3 | 4.41E-16 | 6.92E-16 | 1.02E-15 | <0.37 | 15 | -596 | A | |
| 37 | 05:34:34.11 | 22:02:08.07 | 2.7 | 0.9 | 4.0 | 5.59E-16 | 9.51E-16 | 2.25E-15 | 0.24 | 13 | 350 | C | |
| 38 | 05:34:34.10 | 22:02:12.11 | 2.3 | 0.7 | 3.9 | 6.30E-16 | 1.21E-15 | 2.44E-15 | <0.16 | 13 | 371 | A | |
| 39 | 05:34:33.17 | 22:02:12.99 | 3.7 | 0.8 | 7.3 | 2.65E-16 | 4.90E-16 | 1.95E-15 | <0.21 | 7 | 419 | C | |
| 40 | 05:34:32.38 | 22:02:15.12 | 6.4 | 0.6 | 24.4 | 8.77E-16 | 2.57E-15 | 2.14E-14 | 0.10 | 15 | 188 | C | |
| 41 | 05:34:32.18 | 22:02:18.28 | 1.6 | 0.4 | 2.3 | 4.67E-16 | 7.51E-16 | 1.08E-15 | <0.32 | 11 | | M | |
| 42 | 05:34:33.19 | 22:02:05.94 | 1.0 | 1.0 | 2.6 | 3.00E-16 | 6.07E-16 | 7.86E-16 | 1.16 | 6 | 492 | C | |
| 43 | 05:34:32.99 | 22:01:58.45 | 8.8 | 0.3 | 18.3 | 4.04E-16 | 1.08E-15 | 7.40E-15 | 0.47 | 8 | 646 | C | |
| 44 | 05:34:32.33 | 22:01:45.50 | 4.1 | 0.5 | 16.0 | 6.22E-16 | 1.53E-15 | 9.94E-15 | 0.20 | 10 | 737 | A | |
| 45 | 05:34:31.17 | 22:01:47.03 | 1.0 | 0.7 | 1.3 | 3.28E-16 | 5.22E-16 | 4.23E-16 | <0.67 | 4 | | | $H_2$ next to [O III] circle with [S II] in middle. |
| 46 | 05:34:31.21 | 22:02:00.31 | 2.1 | 0.7 | 6.3 | 6.19E-16 | 1.29E-15 | 3.92E-15 | 0.23 | 19 | -525 | C | multiple spatial components |
| 47 | 05:34:28.64 | 22:02:07.79 | 1.3 | 0.5 | 2.9 | 4.31E-16 | 8.28E-16 | 1.23E-15 | 0.20 | 11 | 201 | C | |
| 48 | 05:34:28.40 | 22:02:11.49 | 2.9 | 0.8 | 7.7 | 4.06E-16 | 8.74E-16 | 3.14E-15 | 0.20 | 12 | 251 | M | |
| 49 | 05:34:28.14 | 22:02:12.48 | 1.4 | 0.4 | 1.9 | 3.28E-16 | 5.44E-16 | 6.23E-16 | <0.32 | 11 | | A | multiple spatial components |
| 50 | 05:34:27.76 | 22:01:49.53 | 2.4 | 0.6 | 5.3 | 4.50E-16 | 8.42E-16 | 2.39E-15 | 0.13 | 10 | 128 | C | |
| 51 | 05:34:27.60 | 22:01:51.56 | 2.3 | 0.7 | 5.5 | 7.33E-16 | 1.83E-15 | 4.05E-15 | 0.10 | 11 | 146 | C | |

| 52 | 05:34:27.08 | 22:01:52.16 | 1.0 | 0.6 | 2.1 | 3.42E-16 | 6.22E-16 | 7.14E-16 | <0.27 | 9 | 121 | A |
| 53 | 05:34:29.28 | 22:00:29.94 | 1.8 | 1.0 | 4.0 | 4.03E-16 | 7.03E-16 | 1.60E-15 | 0.80 | 6 | 696 | C |
| 54 | 05:34:31.02 | 22:00:25.49 | 1.6 | 0.6 | 5.5 | 5.21E-16 | 1.16E-15 | 2.86E-15 | 0.11 | 9 | 774 | C |
| 55 | 05:34:32.88 | 22:00:38.70 | 2.8 | 0.4 | 9.1 | 3.61E-16 | 7.42E-16 | 3.29E-15 | 0.48 | 14 | 789 | M |